\documentstyle[prl,multicol,aps]{revtex}
\input{psfig.tex}
\begin{document}

\draft

\title{Topological entropy and complexity for discrete dynamical systems}

\author{N. Abarenkova}
\address{Centre de Recherches sur les Tr\`es Basses Temp\'eratures,
B.P. 166, F-38042 Grenoble, France\\
Theoretical Physics Department, Sankt Petersburg State University,
Ulyanovskaya 1, 198904 Sankt Petersburg, Russia}

\author{J.-Ch. Angl\`es d'Auriac}
\address{Centre de Recherches sur les Tr\`es Basses Temp\'eratures,
B.P. 166, F-38042 Grenoble, France}

\author{S. Boukraa}
\address{Institut d'A\'eronautique, Universit\'e de Blida, BP 270,
Blida, Algeria}

\author{S. Hassani}
\address{CDTN, Boulevard F.Fanon, 16000 Alger, Algeria}

\author{J.-M. Maillard}
\address{ LPTHE, Tour 16, 1er \'etage, 4 Place Jussieu, 
75252 Paris Cedex, France}

\date{\today}

\maketitle

\begin{abstract}
To test a possible relation between the topological
entropy and the Arnold complexity, and to provide
a non trivial example of a rational dynamical zeta function,
we introduce a two-parameter family of two-dimensional discrete
rational mappings. The generating functions
of the number of fixed points, and 
of the degree of the successive iterates, are both considered.
We conjecture rational expressions with integer coefficients
for these two generating functions.
and  a rational expression for the dynamical
zeta function.
We then deduce algebraic values for the complexity growth and
for the exponential of the topological entropy. 
Moreover, these two numbers happen to be equal
for all the values of the parameters. These conjectures
are supported by a semi-numerical method we explain. This method
also localizes the integrable cases.
\end{abstract}


\begin{multicols}{2}

Chaos is associated to
extreme complexity and ``unpredictability''. Very few
exact results are therefore known, however several exponents
have been introduced to provide measures of the complexity
and further, to classify chaotic systems~\cite{Ott,ASY}. 
The most popular
are the Lyapounov exponents, which have a clear and
intuitive interpretation, but 
require the choice of a metric and of an invariant measure.
They can vary considerably under a very tiny
change of the parameters of the dynamical system
(see~\cite{ASY} p237):
probing fine details, they are not universal.
The same remarks also apply to the metric entropy~\cite{K58,S59}.
By contrast there also exist exponents 
which do not involve any assumption
on the phase space. These exponents, obviously, give a less detailed
description of the system, but are more universal. 
They provide a mean for a classification of dynamical systems.
The topological entropy~\cite{AKM65} $\log{h}$ and $\lambda$ 
the Arnold complexity ~\cite{A} are two examples. 
The topological entropy probes the growth of the number
of stable cycles as a function of their length, and
the Arnold complexity probes the growth of the number of
intersections of a line with its successive iterate.
These two notions give general informations not sensitive
to specific details. 
From an intuitive point of view one can understand the 
importance of the topological entropy, since
the asymptotic behavior of a dynamical
system heavily depends on its fixed points 
and stable cycles: an initial point is often very close
to many basins of attraction, which results in a chaotic motion.
Precisely we introduce the fixed points generating
function $H(t)=\sum_n{h_n} \cdot  t^n $
where $h_n$ is the number of  real, or complex, fixed points 
of the $n$-th power $k^n$ of the mapping. 
The same information is also coded in the so-called
dynamical zeta function $\zeta(t)$~\cite{Ba91,Ru91} 
introduced by Artin and Mazur~\cite{AM65} and related
to the generating function $H(t)$ by
$H(t)  =  t\, \frac{ d}{dt} \ln(\zeta(t))$.
Both functions  only depend on the number of
fixed points, and not on their particular properties
or localization: functions $H(t)$ and $\zeta(t)$
are invariant under topological conjugacy
(see Smale \cite{S67} for this notion). They do not depend on a
specific choice of variable.
$h$, the exponential of the topological entropy characterizes
how the coefficients $h_n$ grow with $n$: $h_n \sim h^n$,
so that $h$ is the modulus of the inverse of the
smallest modulus pole of $H(t)$, if rational.
For a linear dynamic on a torus, the cat map,
the exponential of the topological entropy has been calculated and
found algebraic, i.e. solution of polynomial
with integer coefficients. We are not aware
of other non trivial dynamical systems where 
an algebraic value for the 
exponential of the topological entropy has been calculated.
In this letter we will provide such an example
of a discrete dynamical system with a {\it rational}
dynamical zeta function, and, consequently, algebraic
value for the exponential of the
topological entropy.
This result should not be considered
like a mathematical curiosity : it is similar 
to the rationality of critical exponents in conformal theory.
This algebraicity is the signing
of deeper ``rigid'' structures 
(like Feigenbaum cascades~\cite{F78} are).

In the context of rational mappings it is easy to see that the Arnold 
complexity can be replaced by the complexity growth of the successive
iterations. 
To define the complexity growth $\lambda$
we introduce  the complexity generating function 
$G(t)  =  \sum_n{d_n \cdot t^n}$
where $d_n$ is the degree of any of the numerator, 
or denominator, of the components of the successive iterates 
of the rational mapping under consideration.
When common polynomials factorize
in the numerators and denominators, the coefficients $d_n$
grow slower than expected. We stress that this definition only
apply to rational mappings. 
The complexity growth $\lambda$ characterizes
how coefficients $d_n$ grow with $n$: $d_n \sim \lambda^n$.
Like $h$, the exponential of the topological entropy, complexity 
$\lambda$ is the modulus of the inverse of the
smallest modulus pole of $G(t)$. 
In this letter we claim that the complexity growth $\lambda$
and $h$, the exponential of the topological entropy, are {\it equal}
\begin{equation}
\label{remarkably}
h = \lambda
\end{equation}
This will be tested successfully for a particular class of mappings,
for which both generating functions are conjectured to be rational
and, consequently, the complexity growth
and the exponential of the topological entropy are algebraic. 
We will also give an effective semi-numerical method
to compute these two characteristic numbers.

\begin{figure}
\psfig{file=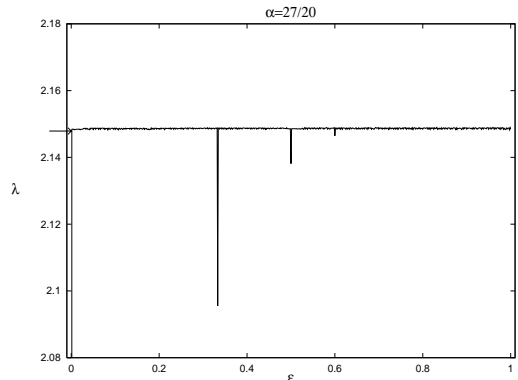,height=5cm,width=7cm}
\caption{The complexity growth $\lambda$ as a function of $\epsilon$
for \\ $\alpha=27/20$. $\epsilon$ is taken of the form $j/720$.
The arrow indicates\\ the conjectured generic value.
\label{fig1}
}
\end{figure}

Let us introduce the discrete rational mapping $k_{\alpha,\epsilon}$
which associates $(u_{n+1},v_{n+1})$ to $(u_{n},v_{n})$
\begin{eqnarray}
\label{uv}
u_{n+1} &=&  1 - u_n + u_n/v_n  \\
v_{n+1} &=& \epsilon + v_n - v_n/u_n 
      + \alpha \cdot (1 - u_n + u_n/v_n)\nonumber 
\end{eqnarray}
This mapping originates from the study of the symmetries of models
of lattice statistical mechanics \cite{BoMaRo95}. Depending
on the actual values of the parameters $\alpha$ and $\epsilon$, the
mapping can have completely different behaviors. 
For example, for $\epsilon=0$ and whatever $\alpha$, as well as
for $\alpha=0$ and  $\epsilon= -1$, 0 , 1/2, 1/3 or 1, the
mapping is integrable, whereas for all other 
values it is not~\cite{BoHaMa97}.
A simple calculation shows that $k_{\alpha,\epsilon}$ is
invertible and that its inverse is also rational. This
property of birationality is of importance in our study.
We have formally calculated the successive powers
of $k_{\alpha,\epsilon}$ for arbitrary $\alpha$ and $\epsilon$,
from which we propose
\begin{equation}
\label{alpeps}
G_{\alpha,\epsilon}(t)  =   \frac{(1+t)^2}{ 1 -  t - 2 t^2 - t^3 }
\end{equation}
The expansion of the conjectured expression Eq.~(\ref{alpeps})
coincides with our results up to the largest power $n=7$ we were able
to compute. Another rational expression 
with the {\it same} denominator is
also obtained if one uses a matricial representation
of the mapping Eq.~(\ref{uv}) \cite{avenir}. 
The expression Eq.~(\ref{alpeps}) of $G_{\alpha,\epsilon}$ 
yields $\lambda \simeq 2.14789$.
To support this conjecture, we have devised a semi-numerical
method to estimate complexity $\lambda$. 
It consists in iterating $k_{\alpha,\epsilon}$ over the field 
of {\it rationals}. During the first steps, some `accidental'
cancellations between numerators and denominators can arise, 
but after this transient regime, the numerators
and denominators get extremely large, and cancellations are only
due to formal simplifications. We then determine how the magnitude
of the four numerators, or denominators, grows with $n$. 
With this method it is possible to raise $k_{\alpha,\epsilon}$
to the 15-th power, and moreover it is easy to scan a large number 
of values of the parameters $\alpha$ and $\epsilon$. 
The calculations are performed with an infinite 
precision C-library \cite{C}. Obviously, this method works
only for rational mappings. On Fig.~\ref{fig1}, one clearly
sees that, for most of the values of $\epsilon$, the complexity
$\lambda$ is extremely close to the expected
value. We call ``specific'' the values of the 
parameters for which the complexity $\lambda$ is
different from 2.14789, they will be discussed later.
We also have formally computed for arbitrary $\alpha$ and $\epsilon$
the fixed points of the powers of $k_{\alpha,\epsilon}$ 
using, once again, the rationality of the mapping. 
This gives 
\begin{eqnarray}
H_{\alpha,\epsilon}(t) &=&   
2 t+2 t^2+11 t^3+18 t^4 \\
 &&  + 47 t^5+95 t^6 + 212 t^7 +  \cdots \nonumber
\end{eqnarray}
From this expression we propose the rational expression for
the generating function of the number of fixed points
$k_{\alpha,\epsilon}$ 
\begin{equation}
\label{cadix}
H_{\alpha,\epsilon}(t) = 
\frac {(2\, + \,3 \,t^2+\, t^3) \cdot  t}{(1-{t}^{2})\cdot(1-t -2 t^2 -t^3)}
\end{equation}
or, equivalently, the dynamical zeta function reads
\begin{equation}
\zeta_{\alpha,\epsilon}(t) = 
\frac{(1+t)(1-t^2)}{(1-t -2 t^2 -t^3)}
\end{equation}
Note that the total number of fixed points of $k_{\alpha,\epsilon}^n$
does {\it not} depend on the actual generic
values of $\alpha$ and $\epsilon$,
however the number of {\it real} fixed points is extremely
dependent on these two parameters. Let us also mention
a {\it local} area preserving property:
the determinant of the Jacobian
of the $n$-th power of $k_{\alpha,\epsilon}$ 
evaluated at each fixed points of $k_{\alpha,\epsilon}^n$ 
is equal to one. The ``visual''complexity of
the phase diagram takes its origin in the real fixed points,
and therefore varies considerably with $\alpha$ and $\epsilon$
\cite{avenir2}.
One sees that the two polynomials giving  exponents
$\lambda$ and $h$ are the same, and consequently we have
the equality $h = \lambda$.
This equality holds for generic values of the parameters,
however, as shown on Fig.~\ref{fig1}, there exist specific
values. These specific values include $\epsilon=1/3, 1/2, 3/5$.
It is then natural to investigate if the equality
of the complexity growth and the exponential of the topological entropy also
holds for the specific values. We have performed calculations
for these values and found that equality (\ref{remarkably})
is always true. The polynomials giving the 
value of $\lambda$ and $h$ are presented in table Tab.~\ref{lespoly}.
Probably, other specific values exist, but they lead to
simplifications occurring at very high orders, and the corresponding
$\lambda$ is too close to the generic value to be
distinguished from it with our method.

\begin{figure}
\psfig{file=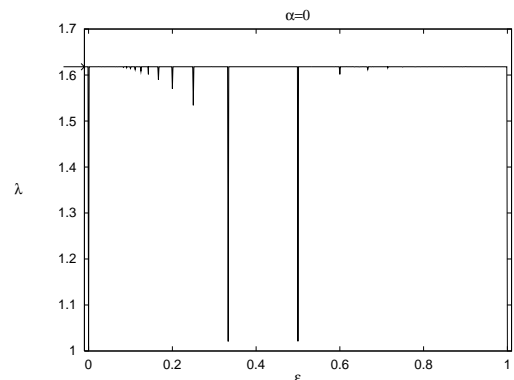,height=5cm,width=7cm}
\caption{The complexity growth $\lambda$ as a function of $\epsilon$
for $\alpha=0$.\\ $\epsilon$ is taken of the form $j/720$, the
values $\alpha=1/7, 1/11, 1/13,5/7$\\ have also been added.
The arrow indicates the conjectured\\  generic value.
\label{fig2}
}
\end{figure}

Besides the specific values mentioned above,
extra simplifications also happen  for $\alpha=0$,
and the complexity is further reduced.
We hence study this special case $\alpha=0$.
In that case, a change of variables~\cite{avenir},
 turns $k_{0,\epsilon}$ into  a  simpler mapping, $k_\epsilon$ 
\begin{eqnarray}
\label{yz}
y_{n+1}  &=&  z_n +1 - \epsilon  \nonumber \\
z_{n+1}  &=&  y_n  \cdot \frac{z_n-\epsilon}{z_n + 1} 
\end{eqnarray}
From now on, the degrees, and the fixed points, 
are those of $k_\epsilon$.
Since the complexity is lower, the semi-numerical method 
presented above is more efficient and it is possible to 
perform calculations beyond the 20-th power. 
The results are displayed on Fig.~\ref{fig2}, where the existence
of integrable values, and non generic values, is clearly seen.
It is simple to see that, if there is no simplification,
$d_{n+1}= d_n + d_{n-1}$ where $d_n$ was introduced in 
the definition of generating function $G(t)$.
In that case 
$G_\epsilon(t)-2 t-1= t \cdot (G_\epsilon(t)-1) + t^2  \cdot G_\epsilon(t)$.
Up to the 20-th power there is no simplification and consequently
we conjecture that, except for the specific values,
the generating function of the complexity growth 
for $k_\epsilon$  is the following rational expression
\begin{equation}
G_\epsilon(t)   =  \frac{1+t}{1 -  t - t^2}
\end{equation}
The corresponding complexity growth is $\lambda \simeq 1.61803$,
in excellent agreement with Fig.~\ref{fig2}.
We have studied the possible equality between $h$ and $\lambda$
for the example $\epsilon=13/25=0.52$. We have chosen this value,
for which  we present a detailed analysis,
because it is generic.
We give in table Tab.~\ref{latable} the number of fixed
points, as well as their properties. The corresponding
phase portrait is very complicated and dominated
by the real fixed points~\cite{avenir2} which are
all saddle or elliptic. We note that the same properties
also holds for the complex fixed points.
The expansion of $H_\epsilon$ can be deduced, up to order eleven,
from Tab.~\ref{latable}. This expansion is compatible with
the very simple rational form for the generating function
of the number of fixed points for
$k_\epsilon$ 
\begin{equation}
H_\epsilon(t)  =  
\frac {(1  + t^2 ) \cdot  t}{(1-t^2 ) \cdot (1-t-t^2)}
\label{lynnie}
\end{equation}
or, equivalently, the dynamical zeta function is
\begin{equation}
\zeta_\epsilon(t)  =  \frac { 1  - t^2 }{1 -t -t^2 }
\label{lyzeta}
\end{equation}
As expected, the two polynomials determining the exponential of the
topological
entropy and the complexity growth are equal, and so are
$\lambda$ and $h$. Both are algebraic numbers.

Coming back to Fig.~\ref{fig2}, we now analyze, for $alpha=0$,
the specific values of $epsilon$.
Let us recall  that $\epsilon=-1,0,1/3,1/2,1$
lead to integrable mappings~\cite{BoHaMa97}. This
corresponds to a polynomial
growth of complexity and of the number of fixed points, 
that is, $\lambda = h = 1$. This is seen on Fig.~\ref{fig2},
except for $\epsilon=-1$, which is out of scale but for which
this is also true. The other specific values are non integrable
and can be partitioned in two sets: $\{1/m;\quad m>3\}$ and
$\{(m-1)/(m+1); \quad m >3\}$. In {\it all} cases the polynomials
giving the complexity growth and the exponential 
of the topological entropy are the
same. These polynomials are listed in Tab.~\ref{lespoly}.

In conclusion, we have given an example
of two-parameter family of
two-dimensional discrete dynamical system
with rational dynamical
zeta function and rational degree generating function $G(t)$.
On this example the exponential of the topological
entropy and the Arnold complexity have the same algebraic value.
A semi-numerical method,
applying to rational transformations only,
has been given, which enables to calculate the complexity
growth, and to localize possible integrable points.
In fact, and this will be detailed
in forthcoming publications, this mapping 
belongs to a ``huge'' family of transformations, for which 
similar results are also obtained.
This family of transformations is so large 
 that (if one believes
in ``some'' universality of dynamical systems) most of the
dynamical systems 
should be very closely ``approximated'' by  transformations 
having  algebraic complexity values.

\acknowledgments{ We thank M. Bellon, P. Lochak, J-P. Marco and 
C. Viallet for discussions, and P.C.E. Stamp for a critical
reading of the manuscript.}

\end{multicols}

\begin{table}
\begin{tabular}{||c|c|c|c|c|c|c|c|c|c|c|c||} 
 $n$                & 1 & 2 & 3 & 4 & 5 & 6 & 7 & 8 & 9 & 10& 11\\ \hline
\# $n$-cycles       & 1 & 0 & 1 & 1 & 2 & 2 & 4 & 5 & 8 & 11& 18\\ \hline
\# elliptic    & 1,0 & 0,0 & 0,0 & 0,0 & 1,0 & 0,0 & 1,0 & 0,0 & 1,0 & 1,3 & 1,0 \\ \hline
\# saddle real      & 0,0 & 0,0 & 1,0 & 1,0 & 1,0 & 0,2 & 1,2 & 1,4 & 3,4 & 1,6 & 5,12 \\ \hline
\# total fixed points  & 1 & 1 & 4 & 5 & 11& 16& 29& 44& 76&121&199\\ 
\end{tabular}
\caption{Number of real (first number) and complex 
(second number) fixed points of $k_{13/25}^n$. 
$n$-cycle means cycle of minimum length $n$. 
\label{latable}
}
\end{table}

\begin{table}
\begin{tabular}{||c|c|c|c|c||}
                                  	&
$\epsilon=1/3$                    	&
$\epsilon=1/2$			  	&
$\epsilon=\frac{1}{m}\quad m>3$   	&
$\epsilon=\frac{m-1}{m+1} \quad m>3$  \\ \hline
$\alpha$ generic                        & 
	$1-t-t^2-2 t^3-t^4-t^5$         &
	$1-t-t^2-2t^3-t^4-2t^5-t^6-t^7$ &
	generic see (\ref{cadix})	& 
	(*)                              \\ \hline
	$\alpha = 0$                    &
 $t$ is $n$-th root of unity            &
$t$ is $n$-th root of unity            &
	$1-t-t^2+t^{m+2}$               & 
	$1-t-t^2-t^{2\, m+1}$              \\
\end{tabular}
\caption{The polynomials giving $\lambda$ 
and $h$ in various specific cases.
The symbol(*) means that $\alpha \neq 0$ and $\epsilon=(m-1)/(m+1)$
are not generic, however the exponents are extremely
close to the generic value, preventing us to compute them reliably.
The case $\alpha \neq 0$ and $\epsilon=1/m$ is generic for $m >3$.
\label{lespoly}
}
\end{table}


\begin{references}
\bibitem{C}
	The multi-precision library gmp is part of the GNU project.
\end{references}
\end{document}